\documentclass{aip-cp}
\usepackage[sort&compress,numbers]{natbib}
\usepackage{rotating}
\usepackage{graphicx}
\usepackage{color}

\begin{document}

\title{First-Principle Investigations of Carrier Multiplication in Si Nanocrystals: a Short Review}

\author[aff1]{Ivan Marri\corref{cor1}}
\author[aff2]{Stefano Ossicini}

\affil[aff1]{CNR-Istituto di Nanoscienze-S3, via Campi 213 A,  41125 Modena, Italy}
\affil[aff2]{Dipartimento di Scienze e Metodi dell'Ingegneria, Universit\'a  di Modena e Reggio Emilia,Via Amendola 2 Pad. Morselli, 42122 Reggio Emilia, Italy and CNR-Istituto di Nanoscienze-S3, via Campi 213 A,  41125 Modena, Italy}
\corresp[cor1]{Corresponding Author: ivan.marri@unimore.it}

\maketitle

\maketitle

\begin{abstract}
Carrier Multiplication (CM) is a Coulomb-driven non-radiative recombination mechanism  which leads to the generation of multiple 
electron-hole pairs after absorption of a single high-energy photon. Recently a new CM process, termed space separated quantum cutting, was introduced to explain a set of new experiments conducted in dense arrays of silicon nanocrystals. The occurrence of this effect was hypothesized  to  generate the formation of Auger unaffected multiexciton configurations constituted by single electron-hole pairs distributed on different interacting naocrystals. In this work we discuss ab-initio  results  obtained by our group in the study of CM effects in systems of strongly interacting silicon nanocrystals.
By solving a set of rate equations, we simulate the time evolution of the number of electron-hole pairs generated in dense arrays of silicon nanocrystals after  absorption of high energy photons,
by describing the circumstances under which CM dynamics can lead to the generation of Auger unaffected multiexciton configurations.
\end{abstract}

\section{INTRODUCTION}
\label{intro}
Nano-structured third generation solar cell devices are promising systems to increase the percentage of electrical energy generated by photovoltaic (PV)
modules. In order to increase  photocurrent production and conversion efficiency we have to maximize the absorption  of the incident solar energy and to minimize  the  relevance of dissipative mechanisms. 
 In single junction solar cells, the maximum theoretical thermodynamic conversion efficiency is defined by the Shockley-Queisser limit \cite{shockley_limit}, that is about the 30\%. This limit move to about 66\% when multijunction solar cells are considered.
The occurrence of non-dissipate recombination mechanisms like the Carrier Multiplication (CM) can reduce the impact of loss thermalization mechanisms on the solar cell performances.
CM is a relaxation mechanism induced by the Coulomb interaction between carriers that leads to the generation of multiple  electron-hole (e-h) pairs after absorption of a single high-energy photon (with an energy at least twice the energy gap of the system). 
CM has been observed in a large number of nanostructured systems, such as PbSe and PbS \cite{ellingson_exp_PbSe_PbS,Semonin_CM,schaller_exp_PbSe_CdSe,trinh_exp_PbSe,nair_exp_PbSe_PbS,schaller_seven}
 CdSe and CdTe \cite{schaller_exp_PbSe_CdSe,schaller_exp_CdSe,gachet_exp_coreshell}, PbTe  \cite{murphy_exp_PbTe},
 InAs \cite{schaller_exp_InAs}, Si  and Ge \cite{beard_exp_Si,Saeed} nanocrystals (NCs).
Recently experimental evidences of new CM dynamics  were observed in Photoluminescence (PL) \cite{timmerman,timmerman_pssa,timmerman_nnano} and in Induced Absorption (IA) experiments \cite{trinh} conducted in low pulse conditions. In the first case, similarities between PL  signals recorded  in  $\textrm{Er}^{3+}$ doped  Si-NCs and in Si-NCs organized in dense arrays  were interpreted by  hypothesizing the occurrence of a new quantum-cutting CM effect, termed space separated quantum cutting (SSQC). When this  effect occurs,  a high energy excited carrier decay toward the band edge (conduction band  (CB) edge for electrons, valence band (VB) edge for holes) by transferring its excess energy to a nearby NC where an extra  e-h pair is generated.   In the second case SSQC was used to interpret 
IA dynamics recorded in low pulse conditions for excitations above and below the CM energy threshold; in this case the intensity recorded for the high excitation photon energy was about two times higher than the one obtained for the lower excitation photon energy, indicating an approximate doubling of the number of generated excitons. Moreover the missing of a fast decay component in the signal recorded for the high energy excitation  leads to exclude the formation of Auger affected multiexciton configurations localized on the same NC \cite{trinh}.
\nonumber
In this work we review results obtained in  ab-initio calculations of CM dynamics in systems of isolated and interacting H-terminated Si-NCs. Ab-initio techniques based on the Density Functional Theory (DFT) have been already applied by our group to calculate structural, electronic, optical and transport properties of  semiconductors of different dimensionality \cite{Iacomino_PRB,iori_doping,govoni_augerbulk,guerra,GUERRA_SM,Ossicini_JNN,DEGOLI2_CRP,iori_NC1,guerra_NC,marri_SSC,PSSC_bert,PSSA_MARRI, PSSB_MARRI}. Here we extend the application of these theoretical tools to
calculate CM lifetimes in systems of isolated and interacting Si-NCs. The calculated CM lifetimes are then introduced as parameters in a set of specific rate equations that are solved to investigate the dynamics of high energy excited states in systems of strongly interacting Si-NCs. Our outcomes are then used to discuss results of Refs. \citealp{timmerman,timmerman_pssa,timmerman_nnano} and  of Ref.  \citealp{trinh}.

\section{METHOD}
\begin{figure}[h!]
  \centerline{\includegraphics[width=\linewidth]{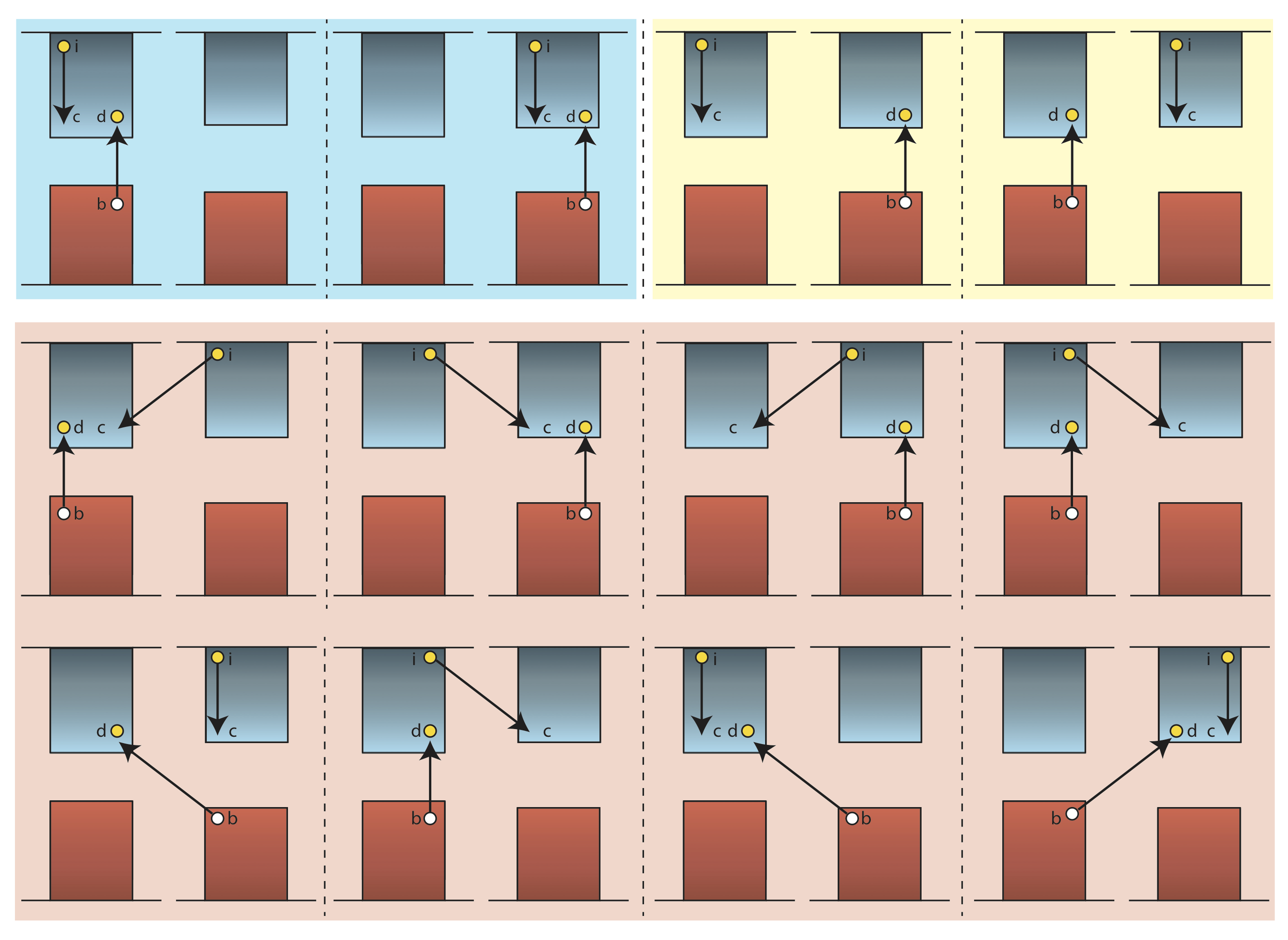}}
  \caption{A schematic representation of one-site CM (blue panels), SSQC (yellow panels) and CDCT (orange panels) processes is reported in the figure. Reprinted with permission from Ref. \citealp{marri_JACS}.}
 \label{Fig1}
\end{figure}

In our approach CM rates are calculated within the DFT by applying first order perturbation theory (Fermi 's Golden rule) to Kohn-Sham (KS) states \cite{allan_stativuoto,delerue_impact1,delerue_impact2,rabani_nanolett_CdSe_InAs}.  CM is therefore described as an impact ionization (II) mechanism (the inverse of the Auger recombination  (AR) process) that follows  the primary photoexcitation event. This scheme permits to correctly estimate CM processes  for bulk systems and nanostructures \cite{califano_apl,califano_CdSe,Franceschetti_PRL,Franceschetti_CMimpact}.
We model the 
decay  of an exciton into a biexciton  as the sum of two processes \cite{delerue_impact1,rabani_nanolett_CdSe_InAs}, one ignited by electron relaxation (decay of an electron in a negative trion, hole is a spectator) and one ignited by hole relaxation (decay of a hole in a positive trion, electron is a spectator). The simultaneous involvement of both particles (the electron and the hole) in the process is neglected in the present treatment. Finally, CM lifetimes are calculated as reciprocal of rates \cite{govoni_nat}.
The CM  rate $R_{n_a, {\bf k}_a}^{e}(E_i)$ for mechanisms ignited by the relaxation of an electron with energy $E_{i}$, is given by:
\begin{eqnarray}
&& R_{n_a, {\bf k}_a}^{e}(E_i)= \sum_{n_c,n_d}^{cond.} \sum_{n_b}^{val.} \sum_{{\bf{k}}_b, {\bf{k}}_c, {\bf{k}}_d}^{1BZ}  4 \pi \Big[ \mid M_D \mid^{2} + \mid M_E \mid^{2} \nonumber \\
&&  + \mid M_D - M_E \mid^{2} \Big] \delta(E_a+E_b-E_c-E_d).
\label{CMe}            
\end{eqnarray}
Similarly, for mechanisms induced by relaxation of a hole with energy $E_{i}$, we have:
\begin{eqnarray}
&& R_{n_a, {\bf{k}}_a}^{h}(E_i)= \sum_{n_c,n_d}^{val.} \sum_{n_b}^{cond.} \sum_{{\bf{k}}_b, {\bf{k}}_c, {\bf{k}}_d}^{1BZ}  4 \pi \Big[ \mid M_D \mid^{2} + \mid M_E \mid^{2} \nonumber \\
&&  + \mid M_D - M_E \mid^{2} \Big] \delta(E_a+E_b-E_c-E_d). 
\label{CMh}            
\end{eqnarray} 
The indexes n and {\bf{k}} identify KS states, 1BZ is the first Brillouin zone and $\mid M_{D} \mid$ and $\mid M_{E} \mid$ are the direct and exchange screened Coulomb matrix elements, respectively. 
In reciprocal space, $\mid M_{D} \mid$ and $\mid M_{E} \mid$ assume the form:
$$
\textrm{M}_D= \frac{1}{V} \sum_{{\bf{G}}, {\bf{G}}'} \rho_{n_d, n_b}({\bf{k}}_d, {\bf{q}}, {\bf{G}}) \textrm{W}_{{\bf{G}} {\bf{G}}'}  \rho_{n_a, n_c}^{\star}({\bf{k}}_a, {\bf{q}}, {\bf{G}'}) \
$$
$$
\textrm{M}_E=
\frac{1}{V} \sum_{{\bf{G}}, {\bf{G}}'} \rho_{n_c, n_b}({\bf{k}}_c, {\bf{q}}, {\bf{G}}) \textrm{W}_{{\bf{G}} {\bf{G}}'}  \rho_{n_a, n_d}^{\star}({\bf{k}}_a, {\bf{q}}, {\bf{G}'}) \
$$
where both ${\bf{k}}_c + {\bf{k}}_d -{\bf{k}}_a -{\bf{k}}_b$ and ${\bf{G}}, {\bf{G}}'$ are vectors of the reciprocal lattice, ${\bf{q}=({\bf{k}}_c -{\bf{k}}_a)}_{1BZ}$ and 
$\rho_{n, m}({\bf{k}}, \bf{q}, {\bf{G}})= \langle n, {\bf{k}} \vert e^{i( {\bf{q}}+ {\bf{G}})\cdot {\bf{r}}} \vert m,  {\bf{k}} - {\bf{q}} \rangle$ is the oscillator strength.
The Fourier transform of the zero-frequency screened interaction $W_{{\bf{G}}, {\bf{G}}'}({\bf{q}}, \omega=0)$ is given by:
\begin{eqnarray}
&&W_{{\bf{G}}, {\bf{G}}'}({\bf{q}}, \omega=0) = v_{{\bf{G}}, {\bf{G}}'}^{bare}({\bf{q}})+W^{p}_{{\bf{G}}, {\bf{G}}'}({\bf{q}}, \omega=0) =\nonumber \\
&& \frac{4 \pi \cdot \delta_{{\bf{G}}, {\bf{G}}'}}{\mid {\bf{q}}+ {\bf{G}} \mid^2} +
\frac{\sqrt{4 \pi e^2}}{\mid {\bf{q}}+ {\bf{G}} \mid} \bar{\chi}_{{\bf{G}}, {\bf{G}}'}({\bf{q}}, \omega=0) \frac{\sqrt{4 \pi e^2}}{\mid {\bf{q}}+ {\bf{G}'} \mid}. \nonumber \\
&&
\label{W}            
\end{eqnarray}
 In Eq. \ref{W}, the first term ($v_{{\bf{G}}, {\bf{G}}'}^{bare}({\bf{q}})$) denotes the bare interaction while the second one ($W^{p}_{{\bf{G}}, {\bf{G}}'}({\bf{q}}, \omega=0)$) includes the screening caused by the medium. These terms are calculated using the many body code Yambo \cite{marini}. $\bar{\chi}_{{\bf{G}}, {\bf{G}}'}({\bf{q}}, \omega)$ is the symmetrized reducible polarizability. Noticeably,  $W^{p}_{{\bf{G}}, {\bf{G}}'}({\bf{q}}, \omega)$ is often given as a function of the  reducible polarization $\chi_{{\bf{G}}, {\bf{G}}'}({\bf{q}}, \omega)$, where: 
$$
\bar{\chi}_{{\bf{G}}, {\bf{G}}'}({\bf{q}}, \omega=0)=\frac{\sqrt{4 \pi e^2}}{\mid {\bf{q}}+ {\bf{G}} \mid}  \chi_{{\bf{G}}, {\bf{G}}'}({\bf{q}}, \omega=0) \frac{\sqrt{4 \pi e^2}}{\mid {\bf{q}}+{\bf{G}}' \mid} .
$$
The reducible polarizability is connected to the irreducible polarizability ${\chi}_{{\bf{G}}, {\bf{G}}'}^0$ by the Dyson equation that, in the Random Phase Approximation (RPA), assumes the form:
$$
{\chi}_{{\bf{G}}, {\bf{G}}'}={\chi}_{{\bf{G}}, {\bf{G}}'}^0 +\sum_{{\bf{G}}_1, {\bf{G}}_2} {\chi}_{{\bf{G}}, {\bf{G}_1}}^0 v_{{\bf{G}_1}, {\bf{G}_2}}{\chi}_{{\bf{G}_2}, {\bf{G}}'}.
$$ 
The presence of off-diagonal elements in the solution of the Dyson equation is related to the inclusion of the local fields that stem from the breakdown of the translational invariance imposed by the lattice.
In our approach the noninteracting response function ${\chi}_{{\bf{G}}, {\bf{G}}'}^0$ is developed in terms of the bare Green's function, that is:
\begin{eqnarray}
&&{\chi}_{{\bf{G}}, {\bf{G}}'}^0({\bf{q}}, \omega)=2\sum_{n,n'} \int_{BZ}\rho_{n', n}^{\star}({\bf{k}}, {\bf{q}}, {\bf{G}})\rho_{n', n}({\bf{k}}, {\bf{q}}, {\bf{G}'}) \times\nonumber \\
&& \left[  \frac{f_{n {\bf{k-q}}}(1-f_{n' {\bf{k}}})}{\omega+\epsilon_{n, {\bf{k}}-{\bf{q}} }  - \epsilon_{n',{\bf{k}} } +i0^+}  -  \frac{f_{n {\bf{k-q}}}(1-f_{n' {\bf{k}}})}{\omega+ \epsilon_{n',{\bf{k}} } -\epsilon_{n, {\bf{k}}-{\bf{q}} }  - i0^+} \right] \nonumber \\
&&
\label{chi}
\end{eqnarray}
where $f_{n {\bf{k}}}$ is the occupation factor of the $\mid n  {\bf{k}}>$ state (zero or one in our case). An exact box-shaped Coulomb cut-off technique is used  in order to remove the spurious Coulomb interaction among replicas (see Ref. \citealp{rozzi}). \\
Systems of non-interacting NCs are simulated by placing a single freestanding NC in an cubic box. In this context H-terminated Si-NCs with diameters ranging from about $1.3$ to $2.4$ nm have been considered. Systems of interacting NCs are instead constructed by placing two NCs in the same simulation box at a tunable separation. In this case CM dynamics are divided in three contributions that are depicted in Fig. \ref{Fig1}. The first one is termed one-site CM and takes into account processes that involve states localized on the same NC (blue panels of Fig. \ref{Fig1}).
The second one is the SSQC (yellow panels of Fig. \ref{Fig1}) that leads to the generation of e-h pairs distributed on different NCs. The third one is the Coulomb Driven Charge Transfer (CDCT, orange  panels of Fig. \ref{Fig1}) that leads to the generation of positive or negative trions. SSQC and CDCT define the so called two-site CM processes.
A detailed quantification of one-site CM, SSQC and CDCT processes is fundamental to interpret the experimental results of Refs. \citealp{timmerman} and \citealp{trinh}.

\section{RESULTS}
\begin{figure}[h!]
\centerline{\includegraphics[width=\linewidth]{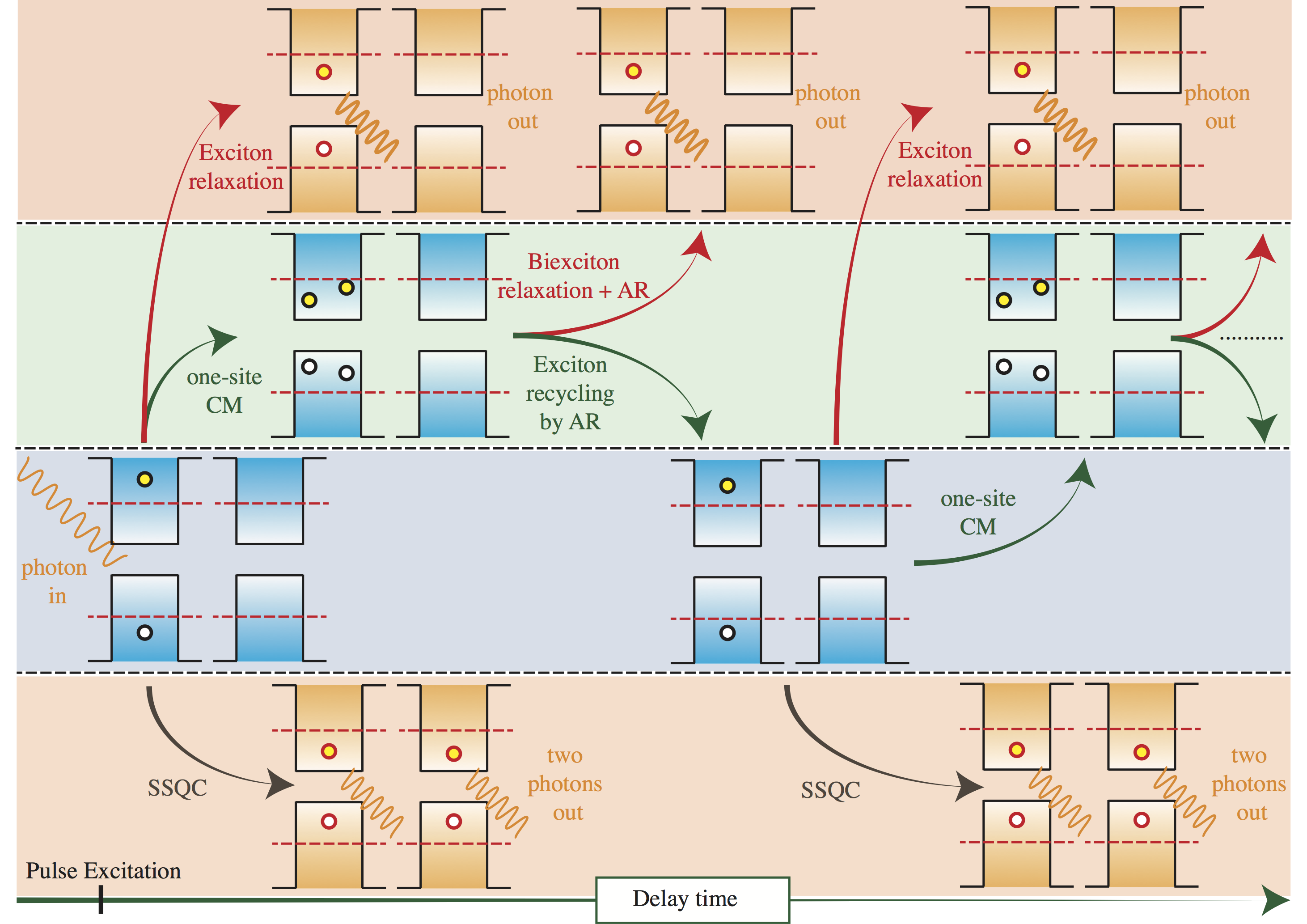}}
\caption{A schematic representation of dynamics represented in Eqs. \ref{rate} is depicted in the figure. An high-energy exciton $\textrm{X}^{\ast}$ generated after pulse excitation can decay by CM into a biexciton XX, by SSQC into two single excitons X+X or can thermalize to X. The biexciton can restore an active configuration $\textrm{X}^{\ast}$, that is an e-h pair that can again decay by CM, through an exciton recycling procedure or can relax and decay to X. $\textrm{X}^{\ast}$ can decay by CM, by SSQC or can relax. The cyclic procedure stops when thermalization mechanisms make impossible to restore an active configuration $\textrm{X}^{\ast}$. Reprinted with permission from Ref. \citealp{marri_JACS}.}
 \label{Fig2}
\end{figure}

\begin{figure}[h!]
  \centerline{\includegraphics[width=\linewidth]{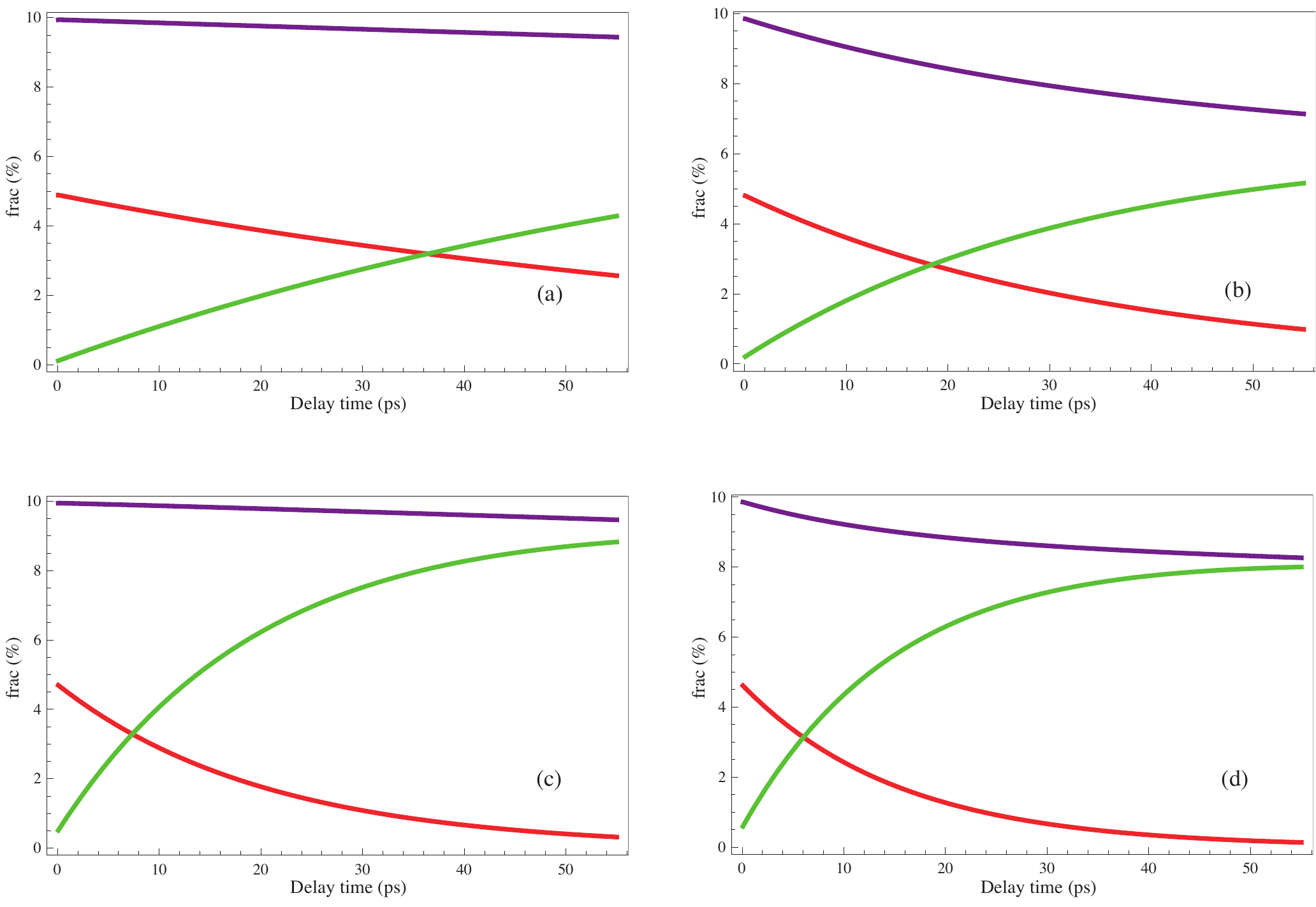}}
  \caption{The solution of Eqs. \ref{rate} are reported in the figure. We assume that only the 5\% of NCs are excited by high energy photons (low pulse conditions). The purple line indicates the total fraction of e-h pairs generated in the sample. Green and red lines represent the  fraction  of biexcitons and single excitons.  
 Calculations have been performed assuming $\tau_{\textrm{one-site}}$ = 0.01 ps, $\tau_{\textrm{Auger}}$ = 1 ps, and $\tau_{\textrm{radiative}}$ = 1000 ps. From panel (a) to panel (d) we have $\tau_{\textrm{SSQC}}$= 1 ps and $\tau_{\textrm{relax}}$ = 5 ps, $\tau_{\textrm{SSQC}}$ = 1 ps and $\tau_{\textrm{relax}}$ = 0.5 ps, $\tau_{\textrm{SSQC}}$ = 0.2 ps and $\tau_{\textrm{relax}}$ = 5 ps, $\tau_{\textrm{SSQC}}$ = 0.2 ps and $\tau_{\textrm{relax}}$ = 0.5 ps. We assume here that the biexciton is always recycled (f = 1). Reprinted with permission from Ref. \citealp{marri_JACS}.}
 \label{rate_eq_1}
\end{figure}

\begin{figure}[h!]
  \centerline{\includegraphics[width=\linewidth]{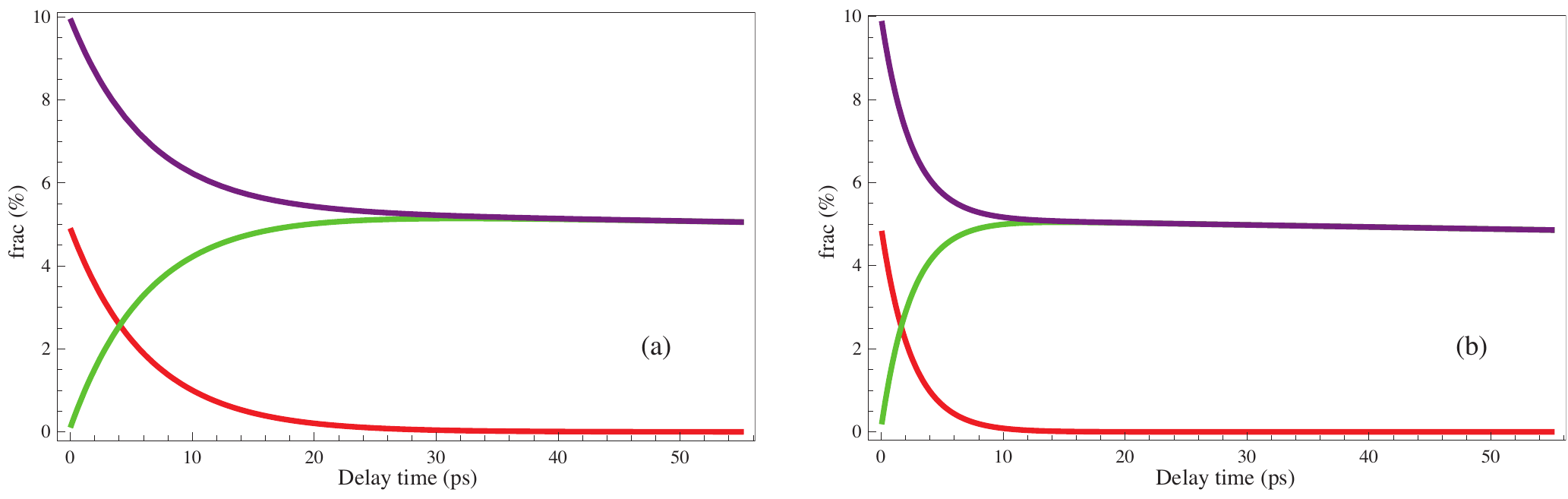}}
  \caption{Solutions of Eqs \ref{rate} are reported in the figure for $f=0.85$ (panel a) and $f=0.60$ (panel b). We assume $\tau_{\textrm{SSQC}}$ = 1 ps and $\tau_{\textrm{relax}}$  = 5 ps. Reprinted with permission from Ref. \citealp{marri_JACS}.}
 \label{rate_eq_2}
\end{figure}

In this section we summarize results obtained by our group in the study of CM processes in Si-NCs. In our works \cite{govoni_nat,marri_JACS,Marri_Beilstein,MARRI_SOLMAT,PSSC_Marri} we have considered four different isolated Si-NCs (the $\textrm{Si}_{35}\textrm{H}_{36}$, $\textrm{Si}_{87}\textrm{H}_{76}$, $\textrm{Si}_{147}\textrm{H}_{100}$ and the $\textrm{Si}_{293}\textrm{H}_{172}$ NCs, with diameters ranging from 1.3 nm to 2.4 nm)  and two systems of interacting Si-NCs obtained by placing two different NCs in the same simulation box at a tunable separation, from 1.0 to 0.4 nm (the  $\textrm{Si}_{147}\textrm{H}_{100} \times \textrm{Si}_{293}\textrm{H}_{172}$ and the $\textrm{Si}_{35}\textrm{H}_{36} \times  \textrm{Si}_{293}\textrm{H}_{172}$).
When first order perturbation theory is adopted to study CM effects in NCs, CM rates and CM lifetimes are given as a function  of   $\textrm{E}_i$. \\
\noindent
Detailed DFT calculations of CM lifetimes in isolated Si-NCs led to conclude that (for more details see Refs. \citealp{govoni_nat} and \citealp{marri_JACS}):
\begin{itemize}
\item CM is active when the initial carrier excess energy, i.e. the energy of the initial carrier  $\textrm{E}_i$ calculated with respect to the band edge (the CB edge for electrons, the VB edge for holes), exceed the energy gap $\textrm{E}_g$. CM lifetimes monotonically decrease with $\textrm{E}_i$ from tens of nanoseconds (near the CM energy threshold) to fractions of femtoseconds (at high energies, that is far from the CM activation threshold).
\item Near the CM energy threshold, CM lifetimes scatter among three orders of magnitude due to the quantum confinement effect. 
\item Far from the activation threshold, CM is proven to be more efficient in Si-NCs than in Si bulk. 
\item When an absolute energy scale is adopted, that is when CM is represented as a function of $\textrm{E}_i$, CM lifetimes seem to be independent upon NC size at high energies.  We have a sort of exact compensation between the effective Coulomb matrix elements, that increase when NC's size decrease, and the density of final states that increase when NC's size increase.
\item When a relative energy scale is adopted, that is when the CM is reported as a function of the ratio between the initial carrier excess energy (energy of the initial carrier calculated with respect the band edge) and  $\textrm{E}_g$, CM is proven to be more efficient in small NCs.
\end{itemize}
By extending our analysis to the study of  systems of interacting Si-NCs, that is the $\textrm{Si}_{147}\textrm{H}_{100} \times \textrm{Si}_{293}\textrm{H}_{172}$  and the  $\textrm{Si}_{35}\textrm{H}_{36} \times \textrm{Si}_{293}\textrm{H}_{172}$, we observed that:
\begin{itemize}
\item when NC-NC separation move from 1.0 nm to 0.4 nm, NCs interplay increase and new CM decay channels are activated.
The delocalization of  wavefunctions over both NCs amplify the importance of two-site CM mechanisms; the relevance of both SSQC and CDCT processes increase when NC-NC separation decrease. 
\item Both SSQC and CDCT lifetimes decrease when $\textrm{E}_i$ increase.
\item When NCs are placed in close proximity, at high values of $\textrm{E}_i$, for the largest and more realistic system (the $\textrm{Si}_{147}\textrm{H}_{100} \times \textrm{Si}_{293}\textrm{H}_{172}$), SSQC lifetimes settle to few ps and CDCT lifetimes settle to fraction of ps.
\item SSQC rate increases when NCs size increase. A similar behavior is not observed neither for one-site CM nor for CDCT.
\end{itemize}
Despite both  SSQC  and CDCT can benefit of the experimental conditions where the presence of an embedding matrix (formation of minibands) or the presence of several interacting NCs (typical condition of three-dimensional realistic systems) are expected to amplify the relevance of two-site CM processes, our results point out that one-site CM mechanisms are always faster than two-site CM processes. In particular a clear  hierarchy of the CM lifetimes $\tau$ emerges from our calculations, that is:
$$
\tau_{\textrm{one-site}} < \tau_{\textrm{CDCT}} < \tau_{\textrm{SSQC}}.
$$
As a consequence, after absorption of a single high energy photon,  a direct generation of single excitons distributed on different interacting NCs is not compatible with our results. Soon after the primary photoexcitation event,  that is the absorption of a high energy photon, we always have the formation of multiexcitons localized in the same NC.
The interpretation of  results obtained in Refs. \citealp{timmerman,timmerman_pssa,timmerman_nnano,trinh} require therefore a more complicated scheme and cannot be ascribed  only to the occurrence of SSQC processes. \\
\noindent
In order to calculate  the time evolution of the number of e-h pairs generated in a sample of strongly interacting Si-NCs after absorption of high energy photons in low pulse conditions, we solve a  set of rate equations where parameters are extracted from our ab-initio simulations. 
The rate equations, represented by the system of Eqs. \ref{rate}, describe the dynamics of Fig.  \ref{Fig2}. 
In this scheme we suppose that  AR can be also an active and not only a destructive mechanism. The final states generated by CM (biexcitons for our NCs) are subject to AR. If AR occurs before the biexciton  thermalization it can be responsible for repopulating high energy levels that lie above the CM energy threshold. The high energy e-h pair generated by AR can therefore, again,  decay  by CM. This new feature of the AR is defined Auger exciton recycling. Noticeably a similar  idea, where AR is considered an active and not only a detrimental effect, was proposed to interpret energy transfer mechanisms in $\textrm{Er}^{3+}$ doped Si-NCs \cite{Navarro-Urrios,Pitanti}. \\
\noindent
\begin{eqnarray}
&&\frac{d}{dt}n_{X^\ast}  =  -\left(\frac{1}{\tau_{\textrm{one-site}}}+\frac{1}{\tau_{\textrm{SSQC}}}+\frac{1}{\tau_{\textrm{relax}}}\right)n_{X^\ast}+\frac{f}{\tau_{\textrm{Auger}}} n_{XX}\vspace{3mm} \nonumber \\
&&\frac{d}{dt} n_{XX} = -\frac{1}{\tau_{\textrm{Auger}}} n_{XX} + \frac{1}{\tau_{\textrm{one-site}}} n_{X^\ast} \vspace{3mm}  \nonumber  \\ 
&&\frac{d}{dt} n_{X} = \left(\frac{2}{\tau_{\textrm{SSQC}}}+\frac{1}{\tau_{\textrm{relax}}}\right)n_{X^*}-\frac{1}{\tau_{\textrm{radiative}}}n_{X} +\frac{1-f}{\tau_{\textrm{Auger}}}n_{XX}
\label{rate}
\end{eqnarray}
\noindent
 We solve the system of  Eqs. \ref{rate} by  assuming an initial configuration that is represented by a high energy e-h pair $\textrm{X}^{\ast}$ localized on a Si-NC (see Fig. \ref{Fig2}). In Eqs. \ref{rate}, $n_{X^\ast}$ is the fraction of above CM threshold excitons $\textrm{X}^\ast$, $n_{X}$ is the fraction of below CM threshold excitons X and $n_{XX}$ is the fraction of biexcitons. 
The e-h pair $\textrm{X}^{\ast}$, generated after absorption of a high energy photon, can thermalize to the conduction and valence band edges  ($\textrm{X}^{\ast} \rightarrow \textrm{X}$), can decay by CM leading to the formation of a biexciton XX ($\textrm{X}^{\ast} \rightarrow \textrm{XX}$) or can decay by SSQC leading to the formation of two e-h  pairs distributed on two different NCs ($\textrm{X}^{\ast} \rightarrow \textrm{X+X}$). While SSQC leads to the formation of space separated e-h pairs (X+X) that can thermalize and recombine radiatively, the biexciton generated by CM can relax to the band edges  or can restore a new configuration that can again decay by CM through an Auger exciton recycling procedure.
The procedure above described,  represented in Fig. \ref{Fig2}, stops when thermalization mechanisms make impossible to generate active configurations that can, again, decay by CM. The scheme of  Fig. \ref{Fig2}, as implemented in Eqs.   \ref{rate}, permits to monitor the time evolution of the number of e-h pairs generated in the sample after absorption of high energy photons (to make the structure of such equations easier, we do not include CDCT mechanisms). 
Moreover, we can  understand how SSQC, AR  and relaxation mechanisms influence such dynamics and we can try to interpret the results obtained in pump and probe experiments conducted on Si-NCs organized in a dense arrays.
\\
In Eqs. \ref{rate} the order of magnitude of  CM one-site, SSQC and  Auger exciton recycling lifetimes ($\tau_{\textrm{one-site}}$,   $\tau_{\textrm{SSQC}}$ and $\tau_{\textrm{Auger}}$, respectively)  have been estimated by first principles calculations. Regarding $\tau_{\textrm{Auger}}$, by using the same Coulomb matrix elements calculated to estimate CM lifetimes, we have proven  that, for NCs of about 2 nm of diameter, $\tau_{\textrm{Auger}}$ settles to about 1 ps. It is important to note that this value is fully compatible with results obtained in experiments conducted on colloidal Si-NCs \cite{beard_exp_Si}. By assuming that biexciton lifetimes $\tau_{biexc}$ scale linearly with NC volume, we can extrapolate from the data reported in Ref. \citealp{beard_exp_Si} the biexciton lifetime for Si-NCs of  2.4 nm of diameter by obtaining  $\tau_{biexc} \approx 6$ ps.
 The  biexciton lifetime represents however an upper limit for the Auger recycling \cite{Note},
as a consequence the result $\tau_{\textrm{Auger}} \approx$ 1 ps is totally reasonable (we note that a biexciton time of 6 ps was also obtained by Klimov et al. \cite{Klimov_science_2} in CdSe colloidal NCs of about 2.4 nm of diameter).
Obviously the present of  surface-related states can inhibit AR processes as pointed out by A. Othonos et al. \cite{Othonos2008}.
\\
The radiative recombination time $\tau_{\textrm{radiative}}$ is assumed to be 1 ns. The parameter $f$ defines the fraction of $\textrm{X}^{\ast}$ that are recycled from XX;  in a realistic system this fraction depends on the competition between biexciton relaxation and Auger recycling rates.
 Remarkably,  due to the complexity of the considered systems, thermalization rates have not been calculated by first principles.
 Starting from results of Ref. \citealp{Boer_nature}, however, cooling time of the order of 1-10 ps are expected for Si-NCs with a diameter of about 2.5 nm. In our approach  we will simulate dynamics of excited carriers by considering both relaxation times of few ps and relaxation time in the sub-ps region.\\
The results of Eqs. \ref{rate} are reported in Fig. \ref{rate_eq_1}  assuming $f=1$  and  an initial population equal to the $5\%$ of $\textrm{X}^\ast$. In the figure the purple line denotes the total fraction of e-h  pairs  generated in the sample after photons absorption. Red and green lines represent the fraction of biexciton (XX) and single excitons (X+$\textrm{X}^{\ast}$) generated in the sample and reported   as a function of the delay time.  The parameters used for the simulations of panels (a)-(d) are reported in the caption of Fig. \ref{rate_eq_1}.
From Fig. \ref{rate_eq_1} we observe that  one-site CM is responsible for probing a double number of excitons with respect to the number of absorbed photons immediately  after the pulse excitation. The time evolution of the number of e-h pairs  is then determined by the combination of one-site CM, SSQC, exciton recycling and thermalization mechanisms. Initially the population is dominated by biexcitons localized in single NCs (red line). Then single excitons localized on different interacting NCs emerge (green line). After the crossing of red and green lines, it is more likely the formation of two excitons in space separated NCs instead of biexcitons localized on single NCs.  Time evolution of the total fraction of e-h  pairs generated after photon absorption  depends on both SSQC and relaxation lifetimes. Anyway the purple lines never show a fast decay component, in agreement with results of Ref. \citealp{trinh}. \\
\noindent
The situation change when only a fraction of biexcitons XX can recycle an active configuration $\textrm{X}^{\ast}$, that is when $f<1$. We have considered two different situations, that is $f=0.85$ and $f=0.60$. Results obtained are reported in Fig. \ref{rate_eq_2}.
It is evident that when thermalization mechanisms reduce  biexcitons energies so that only a fraction of them can restore an active configuration by AR exciton recycling, a fast decay component appears in the purple line. In our model therefore dynamics of excited states are strongly connected with the competition between biexciton relaxation and exciton recycling  mechanisms. When the biexciton thermalization rates are higher than  exciton recycling rates the presence of an Auger  fast decay component  characterize the time evolution of the number of e-h pairs generated  after absorption of high energy photons also  in cases where low pulse conditions are verified.  Here more complicated models have to be proposed in order to explain results of Ref. \citealp{trinh}.

\section{CONCLUSIONS}
In this work we have reviewed some of the results obtained by our group in the study of the CM processes in systems of isolated and interacting Si-NCs. In particular we have shed light on the dependence of the CM dynamics on NCs size and we have summarize the effects induced by NCs interplay on CM dynamics. In order to interpret a set of pump and probe experiments conducted on Si-NCs organized in a dense array we have solved a set of rate equations where the parameters have been determined by first principle calculations. In the implementation of such equations we have re-interpreted the idea of Auger recombination that is  now considered also an active and not only a destructive  process.  This new feature of the Auger has been defined Auger exciton recycling. We have proven that when Auger exciton recycling mechanisms are faster than biexciton relaxation processes our results  can explain experimental evidences.

\section{ACKNOWLEDGMENTS}
The authors thank the Super-Computing Interuniversity Consortium CINECA for support and high-performance computing resources under the Italian Super-Computing Resource Allocation (ISCRA) initiative, PRACE for awarding us access to resource FERMI IBM BGQ, and MARCONI HPC cluser based in Italy at CINECA. Ivan Marri  acknowledges support/funding from European Union H2020-EINFRA-2015-1 programme under grant agreement No. 676598 project "MaX - materials at the exascale".  Stefano Ossicini acknowledges support/funding from Unimore under project "FAR2017INTERDISC".


\nocite{*}
\bibliographystyle{aipnum-cp}%
\bibliography{Manuscript}%

\begin{thebibliography}{51}
\bibitem [{Not()}]{Note}%
  \BibitemOpen
  \href@noop {} {\ }\bibinfo {note} {In our scheme the biexciton decay process
  has to be interpreted as a complicated procedure that occurs thank to the
  combination of recycling and thermalization procedures, and therefore the
  exciton recycling time has to be smaller than the biexciton
  lifetime}\BibitemShut {NoStop}%
\end{thebibliography}

\end{document}